\documentclass[twocolumn,sort&compress]{revtex4-1}
\usepackage[T1]{fontenc}
\usepackage[latin9]{inputenc}
\usepackage{xcolor}
\usepackage{pdfcolmk}
\usepackage{units}
\usepackage{amsmath}
\usepackage{graphicx}
\PassOptionsToPackage{normalem}{ulem}
\usepackage{ulem}

\makeatletter

\providecolor{lyxadded}{rgb}{0,0,1}
\providecolor{lyxdeleted}{rgb}{1,0,0}

\DeclareRobustCommand{\lyxsout}[1]{\ifx\\#1\else\sout{#1}\fi}

\makeatother

\begin{document}

\title{{\normalsize{}Crack initiation in viscoelastic materials}}

\author{{\normalsize{}Yuval Mulla$^{1}$, Giorgio Oliveri$^{2}$, Johannes
T.B. Overvelde$^{2}$, Gijsje H. Koenderink$^{1}$}}
\email{Corresponding author: g.koenderink@amolf.nl}

\affiliation{$^{1}$Living Matter Department, AMOLF, Science Park 104, 1098 XG
Amsterdam}

\affiliation{$^{2}$Designer Matter Department, AMOLF, Science Park 104, 1098
XG Amsterdam}
\begin{abstract}
{\normalsize{}In viscoelastic materials, individually short-lived
bonds collectively result in a mechanical resistance which is long-lived
but finite, as ultimately cracks appear. Here we provide a microscopic
mechanism by which cracks emerge from the nonlinear local bond dynamics.
This mechanism is different from crack initiation in solids, which
is governed by a competition between elastic and adhesion energy.
We provide and numerically verify analytical equations for the dependence
of the critical crack length on the bond kinetics and applied stress.}{\normalsize \par}
\end{abstract}
\maketitle
Liquids cannot fracture, but solids can. We consider the intermediate
case: viscoelastic materials. These materials are made of filaments
or particles interconnected by short-lived bonds. This design theme
of transient networks is commonly used in both natural and man-made
materials such as cytoskeletal polymer networks in cells \citep{Broedersz2010},
physical gels \citep{Dragan2014}, associative and telechelic polymers
\citep{Jasper}, and colloidal gels \citep{Lu2013}.

The molecular dynamics of transient networks lead to interesting macroscopic
mechanics: at times shorter than the bond lifetime the material behaves
like a solid \citep{Broedersz2010}, while on longer time scales the
bonds reorganize and the material deforms viscoelastically \citep{Broedersz2010,Meng2016c,Kroy2012b}.
As a result, transient networks are much more deformable than permanent
networks \citep{Kong2003}. However, viscoelastic materials can resist
mechanical stress only for a limited time, after which the system
suddenly loses its mechanical percolation, a process which is known
as fracturing \citep{Gibaud2010,Sprakel2011,telechelic,Jasper}. This
raises the question how we can design transient networks such that
the robustness against stress is optimized, which requires an understanding
of the mechanism by which transient networks fracture.

Fracturing of viscoelastic materials is often explained by the Griffith
theory of crack initiation in brittle solids \citep{ABR,Jasper,Alvarado2013,Sprakel2009,Lucantonio2015,Gibaud2010}.
The Griffith theory predicts that beyond a critical stress, initial
defects will grow into macroscopic cracks as the elastic energy released
by the crack dominates the surface energy required for separation
\citep{Griffith;1920}. However, this framework was originally developed
for solids, and assumes defects are either static or growing. This
assumption is clearly invalid for viscoelastic materials, as defects
are not static entities but instead continuously appear and heal \citep{Nava2017}.
Therefore, viscoelastic materials require a framework which takes
into account the reversible bond dynamics. 

The seminal work of Bell on cellular adhesion provides such a framework
of reversible bond dynamics under force \citep{Bell1978d}, and has
received considerable attention in studying small-scale systems such
as protein clusters which provide cellular adhesion \citep{Seifert2000,Erdmann2004a},
fracturing of a single colloidal strand \citep{Sprakel2011} and protein
unfolding \citep{Dietz2008}. 

In all of these works, force is assumed to be homogeneously distributed
across all bonds, which appears to be a realistic assumption for microscopically
small systems. Contrarily, in the context of viscoelastic materials,
theoretical \citep{Head2003,Colombo2013} and experimental work on
both synthetic gels \citep{Basu2011} and biopolymer networks \citep{Liu2007}
has revealed non-affine deformations upon application of global stress
\citep{Wen2012a}. Indeed, imaging of various networks under stress
showed inhomogeneities of the local force which are strongly correlated
in space \citep{Majmudar2005,Guo2014,Arevalo2015}. These inhomogeneities
might be negligible when considering bulk properties such as the average
bond lifetime under stress \citep{Vaca2015,Gralka2015}, but likely
play a key role in crack initiation. In situ imaging of stressed viscoelastic
materials revealed that fracturing occurs via well-defined cracks
\citep{Gladden2007,ABR,Jasper,Foyart2016} rather than via the homogeneous
degradation expected from the Bell model \citep{Seifert2000,Erdmann2004a,Dietz2008,Novikova2013,Bell1978d,Vaca2015,Gralka2015},
suggesting local rather than global load sharing. At which length
scale does the global load sharing assumption become inaccurate? And
what determines the fracturing properties of a system of reversible
bonds under load beyond this length scale?

\begin{figure}
\includegraphics[width=1\columnwidth]{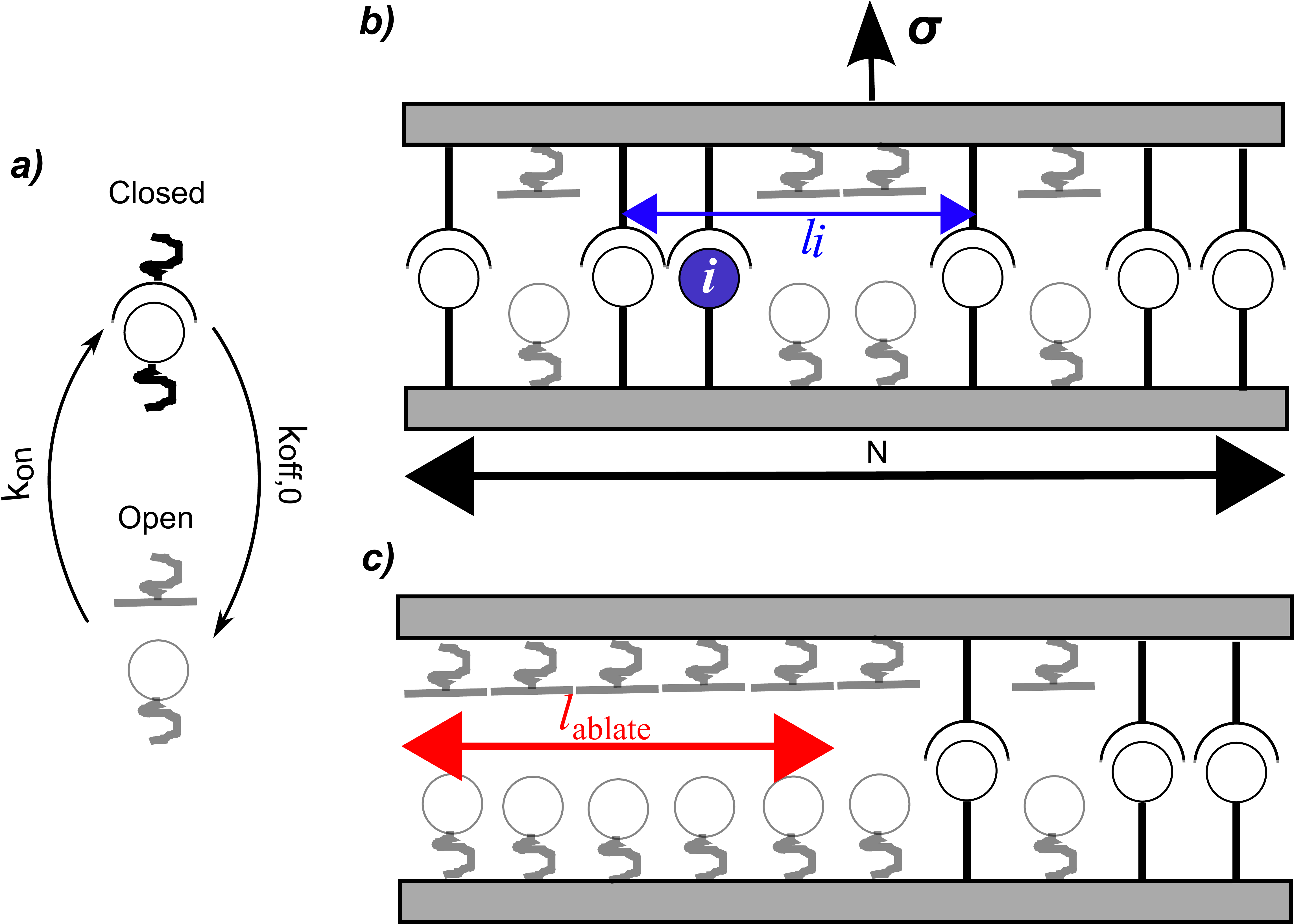}

\caption{\textbf{Schematic of model}{\small{} a) Bonds switch from an open
to a closed state with rate $k_{\textrm{on}}$ and reverse with a
rate of $k_{\textrm{off,0}}$ in the absence of force. b) Bonds in
a closed state share an applied load $\sigma$, where the load distribution
depends on the distance $l_{i}$ from bond $i$ to its nearest neighbors.
c) We perform bond ablation experiments by opening all bonds in $l_{\textrm{ablate}}$
adjacent positions to investigate the critical length required for
triggering fracturing. Periodic boundary conditions are used to prevent
edge effects from influencing the results. }\textbf{\label{fig:Model-of-load}}}
\end{figure}

To answer these questions we developed a minimal model that includes
reversible bond dynamics (figure \ref{fig:Model-of-load}a) in the
simplest possible 'material' that is capable of exhibiting spatial
inhomogeneity required for studying crack initiation: bonds distributed
over a 1D-space, subject to mechanical stress. To account for inhomogeneous
load sharing, we assume a force distribution that depends on the local
bond spacing (figure \ref{fig:Model-of-load}b). We show that this
minimal model system exhibits spontaneous crack initiation and subsequent
fracture, in a manner that is consistent with experimental observations
in wide range of viscoelastic materials \citep{Gladden2007,ABR,Jasper,Foyart2016}.
We verify our results by comparison with a mechanical model. We study
the process of crack initiation in more detail by locally 'ablating'
bonds (figure \ref{fig:Model-of-load}c), which reveals a critical
crack length beyond which fracturing occurs. We provide analytical
equations describing the process of crack initiation on basis of the
nonlinear bond dynamics, and predict the dependence of the critical
crack length on both bond properties and applied stress. Our work
reveals that the process of crack initiation in viscoelastic materials
is fundamentally different from that in traditional solids, as a consequence
of the reversible bond dynamics.

We initialize a one-dimensional (1D) network with $N$ equally spaced
binding sites using periodic boundary conditions, each bond having
a probability $K$ to start in a closed state. Next we model the dynamics
of the bonds with a kinetic Monte Carlo scheme \citep{Gillespie1976}
using the following bond dynamics:

\begin{equation}
K=\frac{k_{\textrm{on}}}{k_{\textrm{on}}+k_{\textrm{off,0}}}
\end{equation}
where $k_{\textrm{on}}$ is the rate of bond closing and $k_{\textrm{off,0}}$
the rate of bond opening in the absence of force (figure \ref{fig:Model-of-load}a).
We normalize time by the on-rate, $k_{\textrm{on}}$. The off-rate
increases exponentially with the applied force $f$ on the bond in
keeping with the Bell model \citep{Bell1978d}:

\begin{equation}
k_{\textrm{off}}(f_{i})=k_{\textrm{off,0}}\cdot\exp(\frac{f_{i}}{f_{\textrm{1/e}}})\label{eq:off-rate}
\end{equation}
where $f_{\textrm{\textrm{1/e}}}$ is the force where the off-rate
has fallen to 1/e of $k_{\textrm{off,0}}$. We calculate the force
per bond $f_{i}$ via

\begin{equation}
f_{i}=\alpha_{i}\cdot\sigma
\end{equation}
where $\sigma$ is the stress on the system and $\alpha$ is a yet
to be defined stress intensity factor per bond. In global load sharing,
the applied stress is equally divided over all bonds. To investigate
the effect of inhomogeneous force distribution as present in any network
under stress \citep{Majmudar2005,Guo2014,Arevalo2015}, we investigate
a local load sharing model. In this model, we assume that the force
distribution is dependent on the distance $l_{i}$ of a bond to its
nearest neighbor on both sides (figure \ref{fig:Model-of-load}b).
Explicitly, we define a stress intensity factor $\alpha$ on a closed
bond at site $i$ by:

\begin{equation}
\alpha_{i}=\begin{cases}
N\cdot\frac{l_{i}}{\Sigma_{i}l_{i}} & Local\\
\frac{N}{\Sigma_{i}n_{i}} & Global
\end{cases}\label{eq:load_sharing}
\end{equation}
where $n_{i}$ equals 1 when the bond is closed and 0 when the bond
is open. Note that in both modes of load sharing the total amount
of force is independent of the bound fraction and normalized by the
system size,$\mbox{\ensuremath{\frac{\sum_{i}f_{i}}{N}}=\ensuremath{\sigma}}$.
We normalize the applied stress by the bond force sensitivity $f_{\textrm{1/e}}$
. After calculating the force on all bonds, we employ a kinetic Monte
Carlo step to either open or close a bond stochastically. We repeat
this process of stochastic bond removal/addition until all bonds are
removed.

\begin{figure}
\includegraphics[width=0.5\columnwidth]{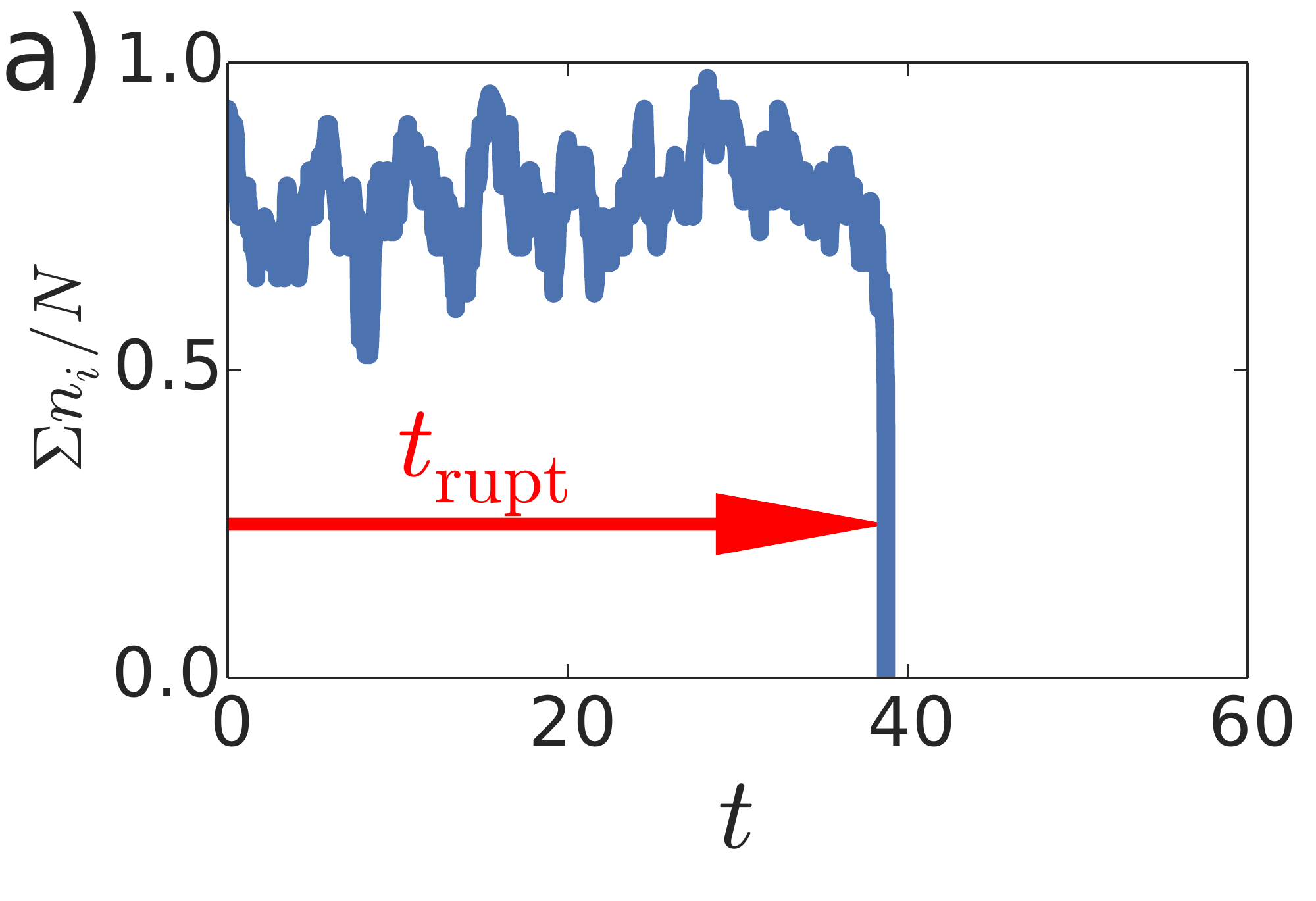}\includegraphics[width=0.5\columnwidth]{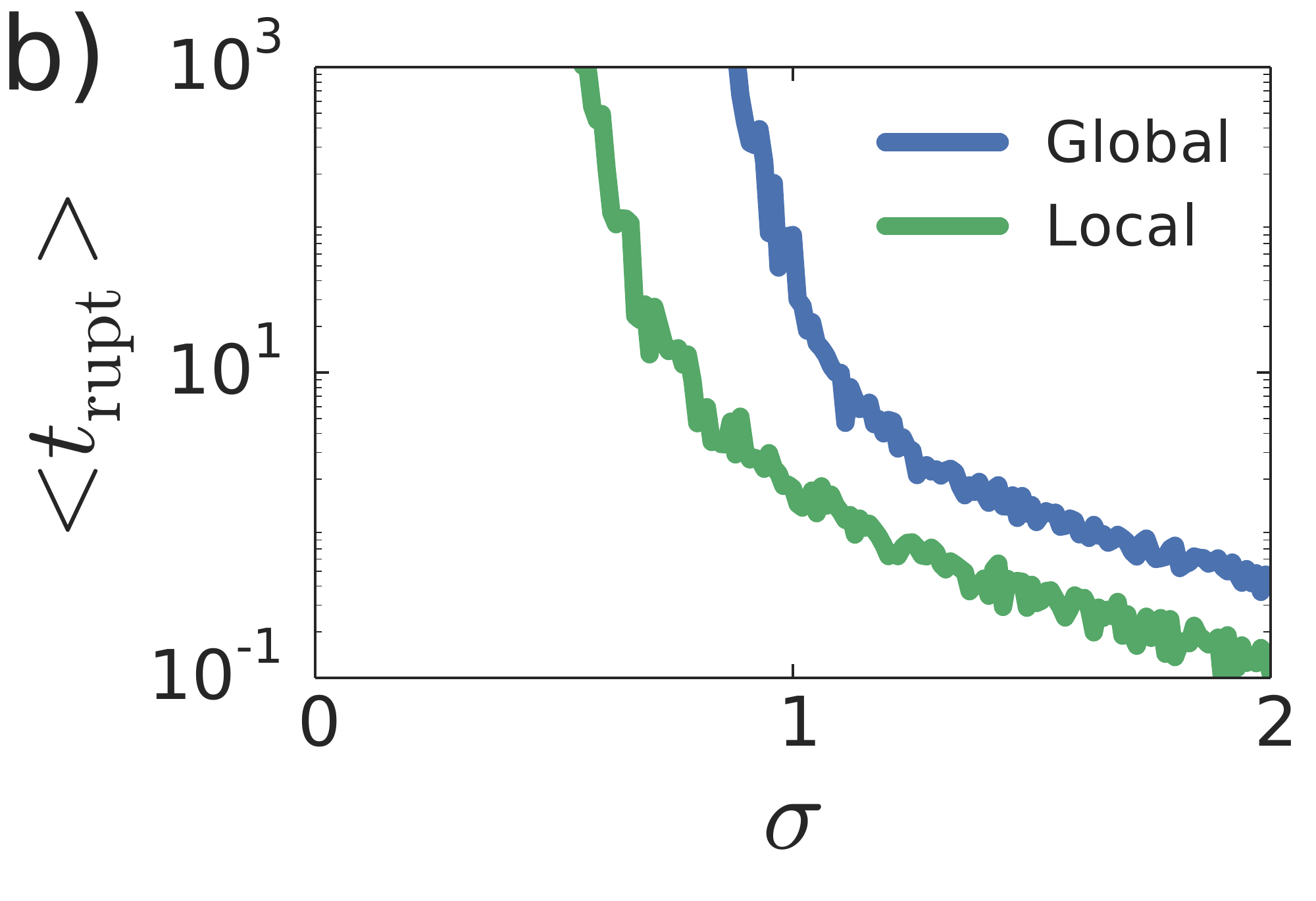}

\includegraphics[width=0.5\columnwidth]{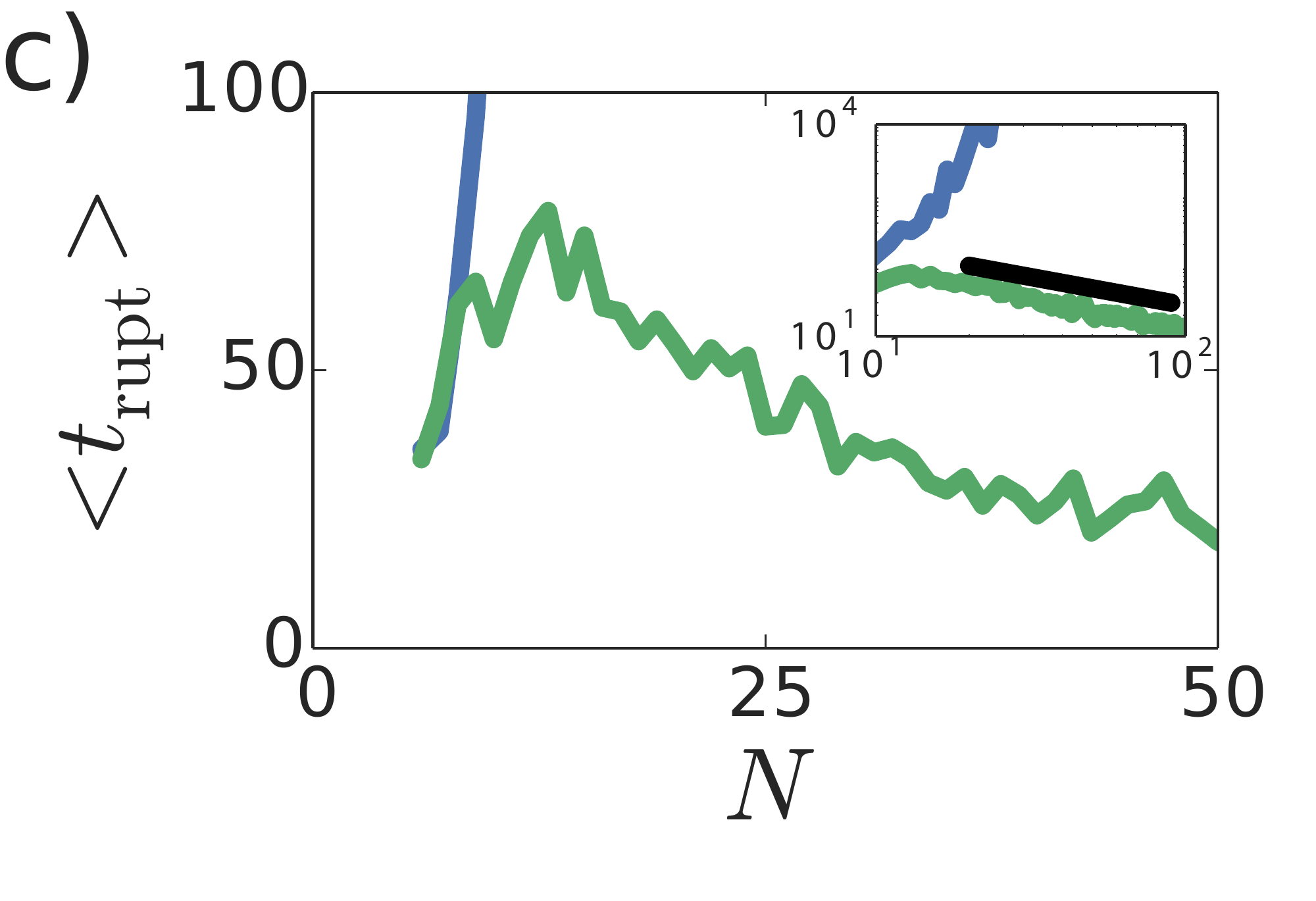}\includegraphics[width=0.5\columnwidth]{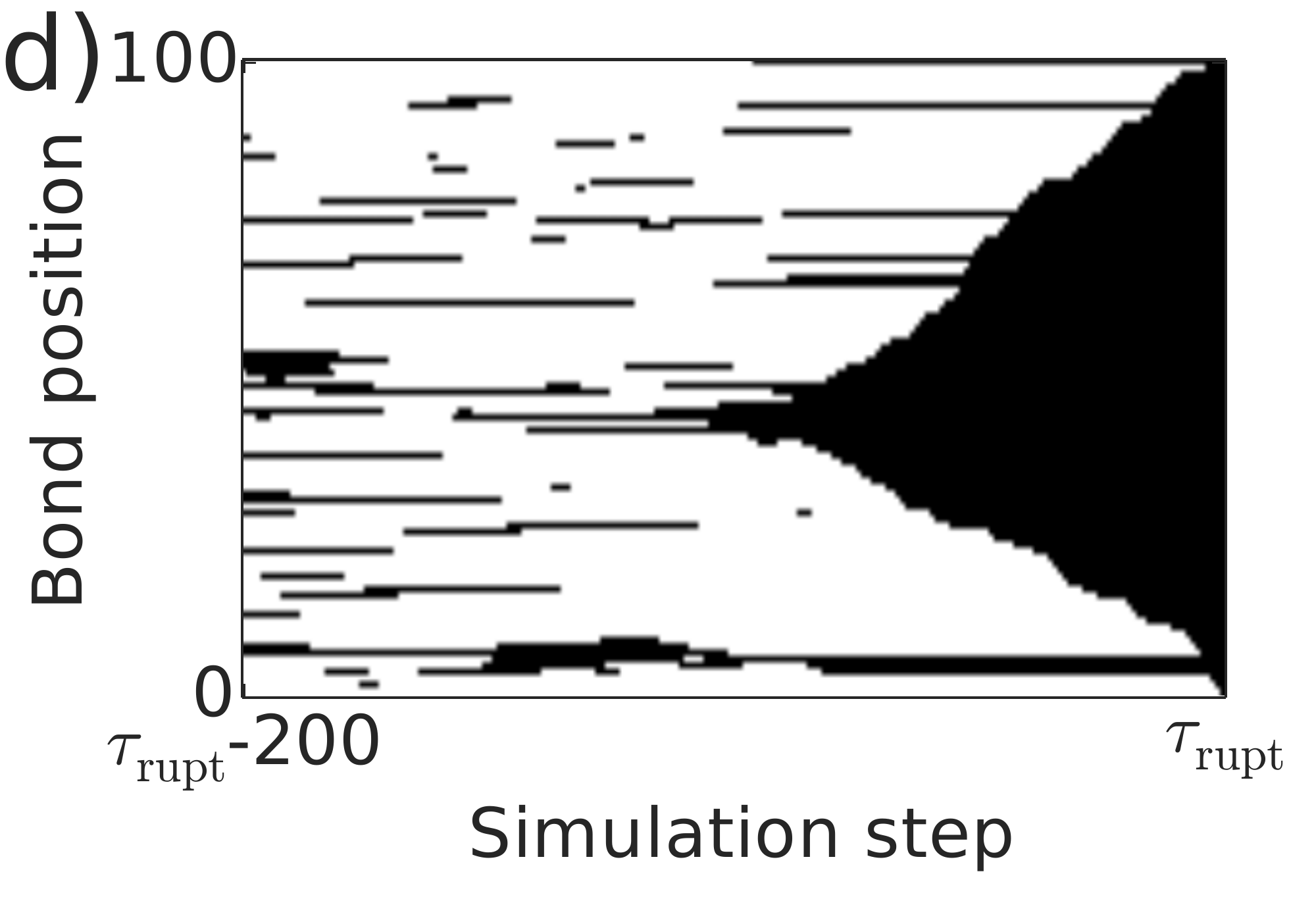}

\caption{\textbf{Stochastic rupture of simulated 1D transient networks subject
to a mechanical stress.}{\small{} a) Typical example of the fraction
of closed bonds in time upon application of stress, after $t=t_{\textrm{rupt}}$,
spontaneous fracture occurs ($K=0.9$, $N=20$, $\sigma=0.7$, global
load sharing). b) Stress dependence of rupture time. Although quantitatively
different, global and local load sharing show qualitatively similar
behavior with two exponential regimes with a cross over at around
$\mbox{<\ensuremath{t_{\textrm{rupt}}}>\ensuremath{\approx}1}$ ($K=0.9$,
$N=20$). c) The system size dependence of the rupture time reveals
a qualitative difference between global and local load sharing: whereas
the rupture time increases with system size for global load sharing,
local load sharing shows an optimum in strength at a well-defined
system size ($K=0.9$, $\sigma=0.7$). Inset: same data on a log-log
scale, showing that after a critical system size $\mbox{<\ensuremath{t_{\textrm{rupt}}}>\ensuremath{\sim N^{-1}}}$
for local load sharing. d) }Kymograph of crack initiation under local
load sharing (white=closed, black=open).\textbf{ }Plotting the bond
state as a function of position (y-axis) versus simulation step (x-axis)
clearly reveals how bond opening proceeds via a well-defined crack.
The x-axis shows simulation step rather than time, as the crack propagation
is orders of magnitude faster than the crack initiation (figure \ref{fig:FEM-and-local}a).{\small{}
\label{fig:FEM-and-local}}}
\end{figure}

As shown in figure \ref{fig:FEM-and-local}a, the fraction of closed
bonds fluctuates over time, until it drops precipitously to zero at
a certain moment that we denote as the rupture time. The rupture time
is exponentially distributed - indicative of the stochastic nature
of fracturing \citep{SI}. To test the sensitivity of the average
rupture time to the applied stress, we perform simulations at a fixed
network size and fixed bond affinity for different levels of stress
($N=20$, $K=0.9$, $\sigma=0.5...2$). For both local and global
load sharing, we find that the average rupture time shows two distinct
regimes with a transition around $\mbox{<\ensuremath{t_{\textrm{rupt}}}>\ensuremath{\approx}1}$
(figure \ref{fig:FEM-and-local}b). As we will explain later on, these
regimes correspond to a metastable network at low stress and an unstable
network at high stress. Importantly, networks with local load sharing
are markedly less robust than globally load sharing networks, with
smaller average rupture times at all stresses. 

To test how the system size influences the average rupture time, we
perform simulations for networks with $N$ varying between 5 and 100
(figure \ref{fig:FEM-and-local}c). In case of global load sharing,
we see that the average rupture time monotonically increases with
system size, as the relative fluctuations of the fraction of closed
bonds ($\frac{\Sigma_{i}n_{i}(t)}{N}$) decreases \citep{Bell1978d,Seifert2000,Erdmann2004a,Sprakel2011}.
In case of local load sharing, we find similar rupture times as compared
to globally load sharing networks for small $N$. But strikingly,
beyond a critical length (around $N=12$ for these conditions), we
find that only in case of local load sharing the rupture time decreases
with increasing system size, according to $\mbox{<\ensuremath{t_{\textrm{rupt}}}>\ensuremath{\sim N^{-1}}}$
(inset of figure \ref{fig:FEM-and-local}c). This dependence suggests
a constant crack initiation rate for every 12 bonds at this particular
stress. Indeed, kymographs of simulations using local load sharing
reveal that fracturing proceeds via cracks rather than homogeneous
degradation (figure \ref{fig:FEM-and-local}d).

\begin{figure}
\includegraphics[width=0.5\columnwidth]{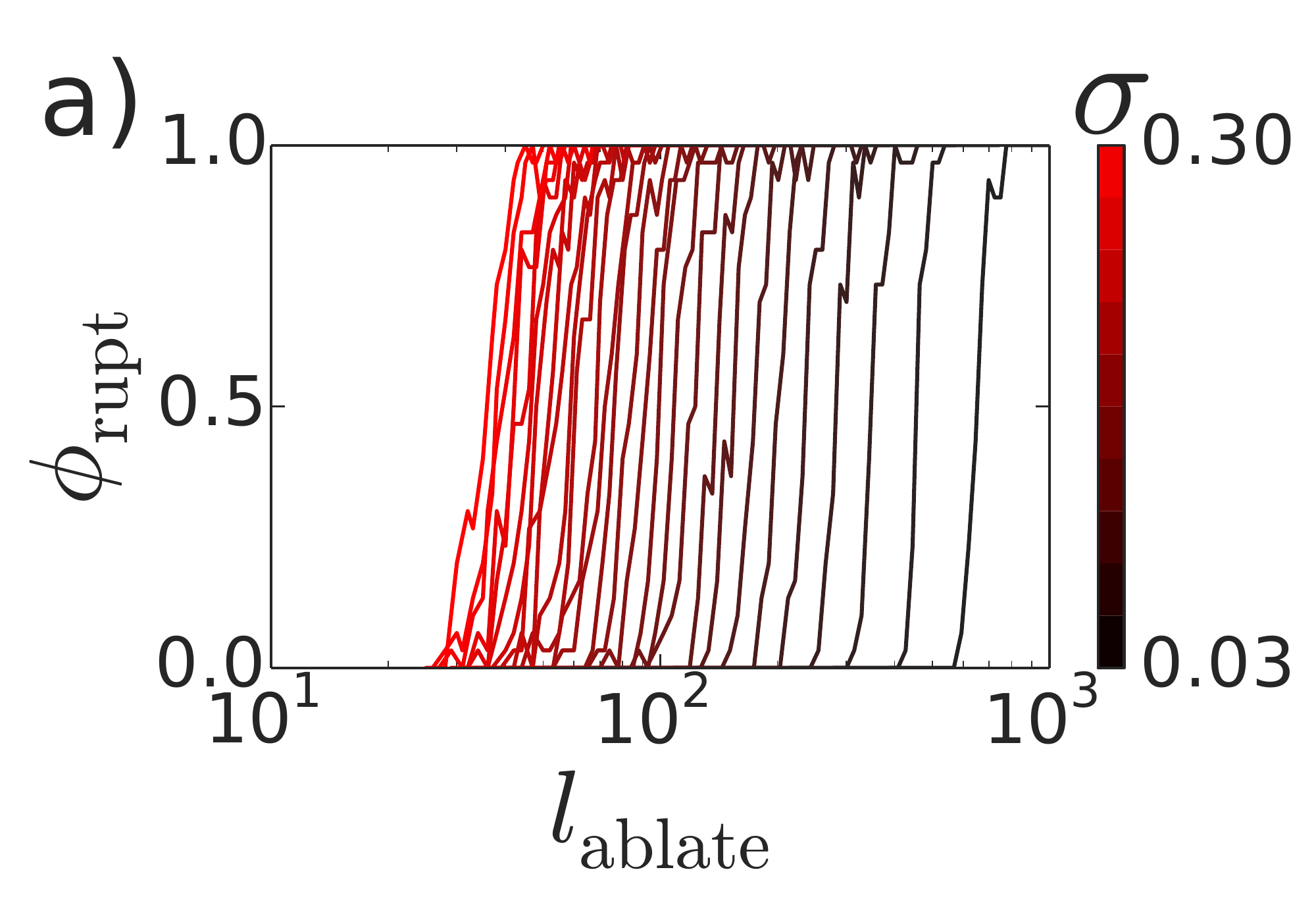}\includegraphics[width=0.5\columnwidth]{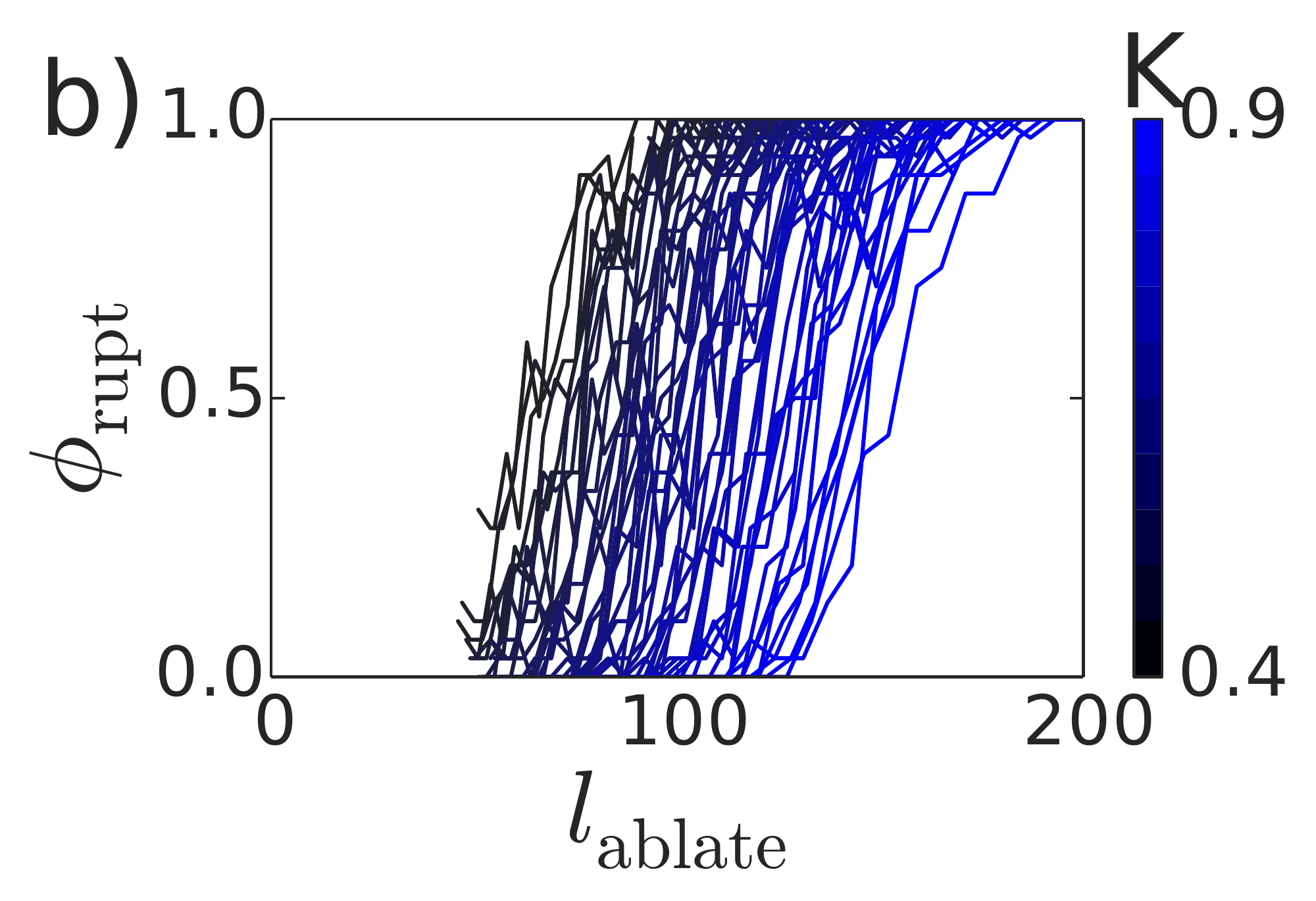}

\caption{\textbf{Characterization of critical crack length in local load sharing}
{\small{}Ablation experiments were performed by first equilibrating
the system under stress until $t=1$, next bonds were ablated: $n_{i}=0$
for $\mbox{i=0...\ensuremath{l_{\textrm{ablate}}}}$. After the ablation,
the network was studied up to $t=2$. This experiment was repeated
30x per condition, and the fraction of observed ruptures $\phi_{\textrm{rupt}}$
was recorded. We plot the ablation size $l_{\textrm{ablate}}$ versus
the fraction of observed ruptures $\phi_{\textrm{rupt}}$ for different
values of a) applied stress $\sigma$ at $K=0.9$ or b) bond affinity
$K$ at $\sigma=0.1$. Control ablation experiments using the same
parameter with global load sharing never showed fracturing.\label{fig:Characterization-of-hidden}}}
\end{figure}

To understand what sets this critical length for crack initiation,
we performed 'ablation experiments' (figure \ref{fig:Model-of-load}c):
first we equilibrate the network under stress, next we remove all
bonds in $l_{\textrm{ablate}}$ adjacent positions, then we study
whether bond ablation triggered network fracturing. We chose the system
size $\mbox{N=\ensuremath{l_{\textrm{ablate}}\cdot}10}$, such that
the system is large compared to the number of ablated bonds, yet small
enough to allow for equilibration without spontaneous crack initiaton.
Figure \ref{fig:Characterization-of-hidden}a shows how fracturing
becomes more likely upon increasing the ablation size $l_{\textrm{ablate}}$,
and that the required ablation size $l_{\textrm{ablate}}$ to initiate
fracturing decreases with the applied stress $\sigma$. Figure \ref{fig:Characterization-of-hidden}b
shows that an increase of bond affinity $K$ increases the critical
ablation size $l_{\textrm{ablate}}$ required for triggering fracture.

To quantitatively understand the ablation data, we define crack length
$L$ as the largest bond distance $l_{i}$ in the system. In case
of global load sharing, the force on bonds at the edge of the crack
stays independent of $L$ as long as $\mbox{L\ensuremath{\ll}N}$,
so ablation does not induce fracture. By contrast, in case of local
load sharing, the force on the bond at the edge of the crack is $\mbox{f=\ensuremath{\sigma\cdot}L\ensuremath{\cdot\nicefrac{1}{2}}}$
(the factor $\nicefrac{1}{2}$ is because the load on the hole is
shared by the bonds at both ends). Thus, $k_{\textrm{off}}$ exponentially
grows with the crack size due to a linearly increasing force, whereas
the chance of rebinding increases only linearly due to a larger area
in which rebinding can occur. As a result, for large enough $l_{\textrm{ablate}}$,
bond unzipping will occur for any system under stress. 

\begin{figure}
\includegraphics[width=0.33\columnwidth]{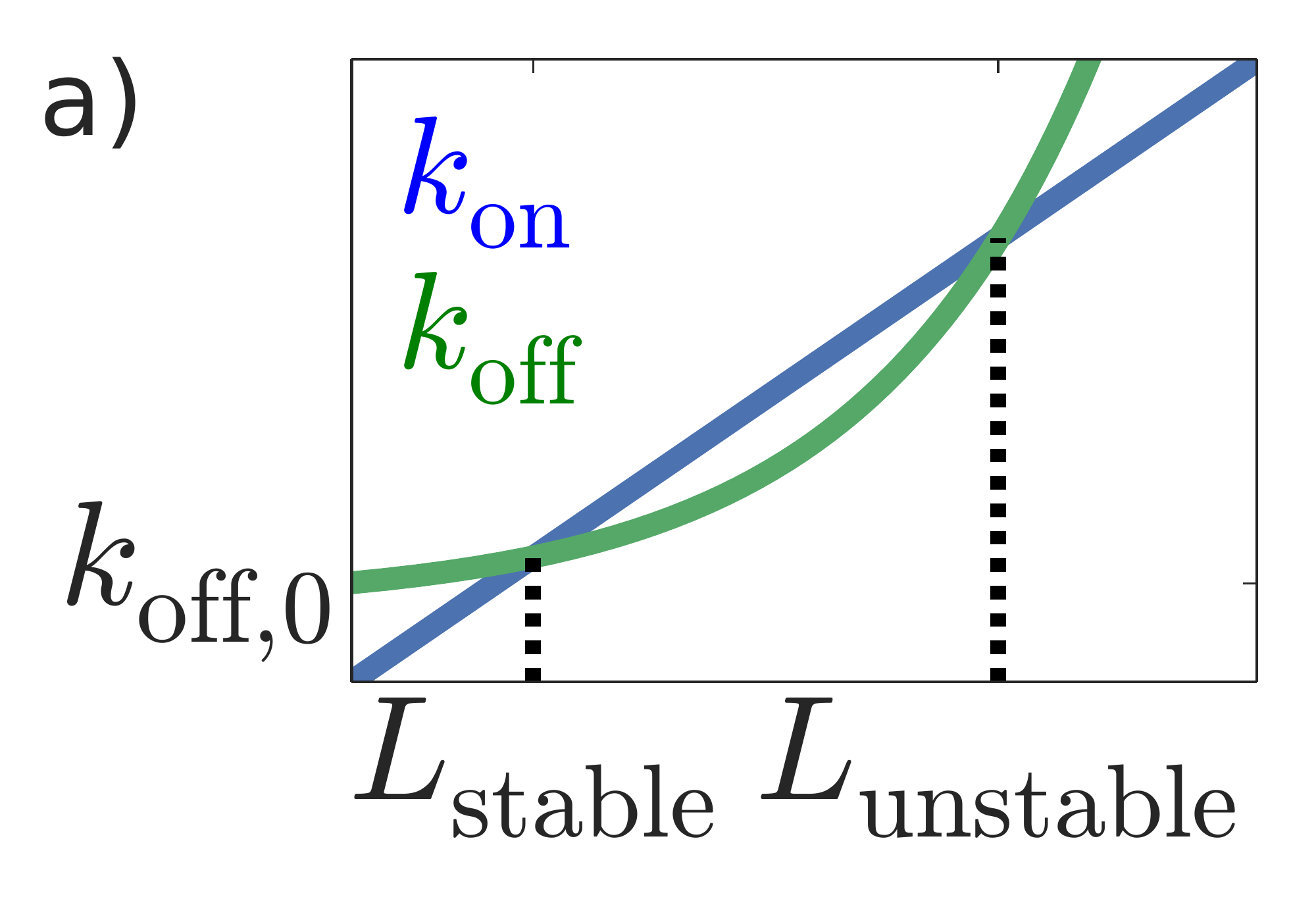}\includegraphics[width=0.33\columnwidth]{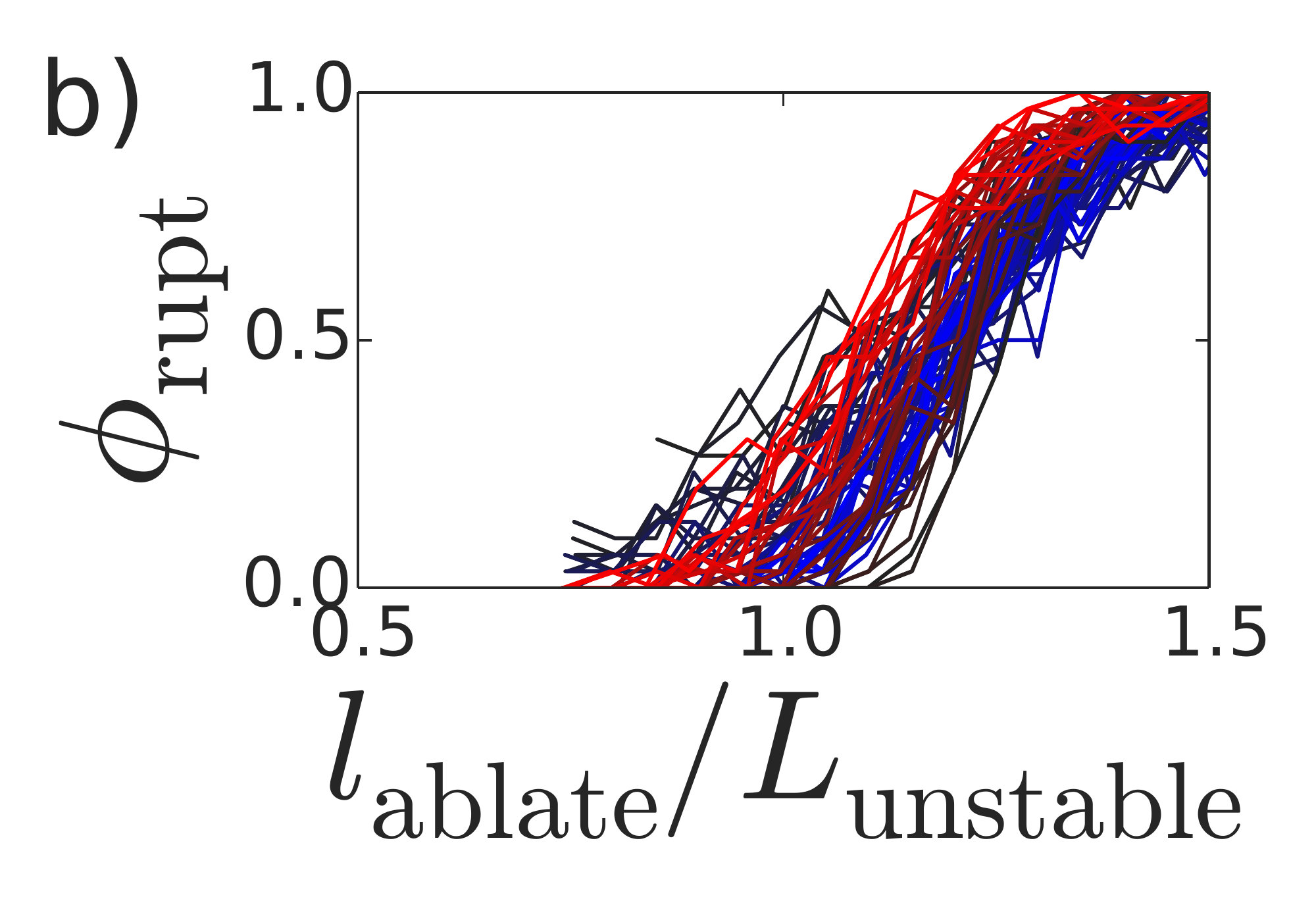}\includegraphics[width=0.33\columnwidth]{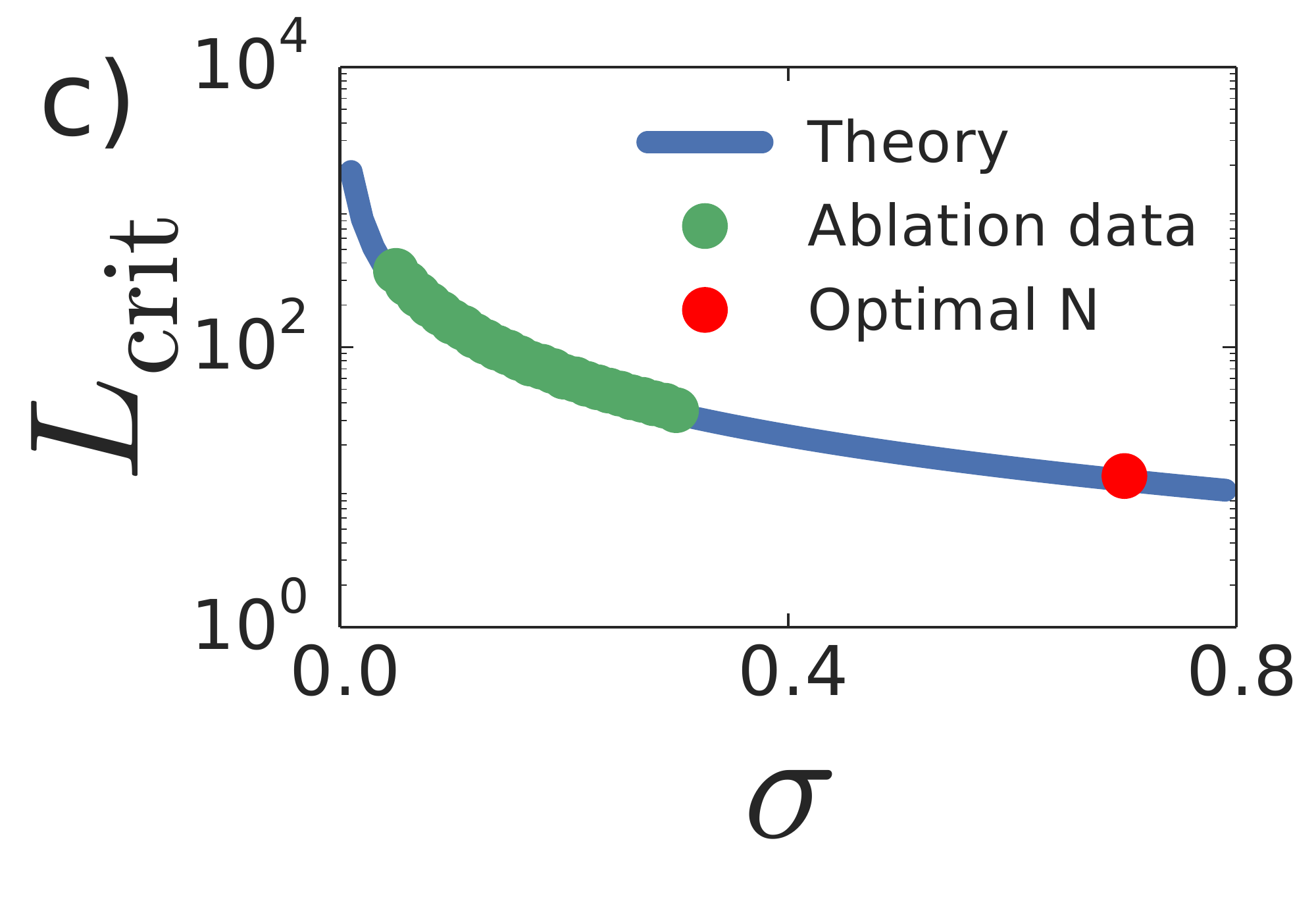}

\caption{\textbf{\small{}Comparison between theory and simulation}{\small{}
a) As a function of crack size, the on-rate increases linearly, whereas
the off-rate increase exponentially. As a result, the crack becomes
unstable after $L_{\textrm{unstable}}$. b) All data from figure \ref{fig:Characterization-of-hidden}
can be collapsed obtained onto a single master curve by normalizing
according to $l_{\textrm{ablate}}/(L_{\textrm{unstable}}(\sigma,K)$).
c) Equation \ref{eq:tipping_point} quantitatively predicts both the
critical length for ablation, and the width at which the maximal rupture
time is observed in figure \ref{fig:FEM-and-local}.}\label{fig:theory_collapse}}
\end{figure}

We are interested in the length $L_{\textrm{unstable}}$ at which
the crack becomes unstable. As a first order approximation, we can
find the fixed points of crack length $\overset{*}{L}$ by calculating
the length at which the rates of bond opening and closing are equal
(figure \ref{fig:theory_collapse}a):
\begin{equation}
2\cdot k_{\textrm{off}}(\nicefrac{1}{2}\cdot\sigma\cdot\overset{*}{L})\approx k_{\textrm{on}}\cdot\overset{*}{L}\label{eq:binding-unbinding}
\end{equation}
This condition is met at the average bond distance at equilibrium,
$L_{\textrm{stable}}$, and the bond distance at the unstable point,
$L_{\textrm{unstable}}$: 
\begin{equation}
L_{\textrm{stable}}(\sigma,K)\approx2\cdot\frac{W_{0}(-\sigma\cdot(\frac{1}{K}-1))}{-\sigma}\label{eq:steady_point}
\end{equation}
\begin{equation}
L_{\textrm{unstable}}(\sigma,K)\approx2\cdot\frac{W_{-1}(-\sigma\cdot(\frac{1}{K}-1))}{-\sigma}\label{eq:tipping_point}
\end{equation}
where $W$ is the Lambert W function with $W_{0}$ the main branch
and $W_{-1}$ the second branch \citep{LambertW}. Note that the network
transitions from metastable to unstable at $\mbox{\ensuremath{L_{\textrm{stable}}}=\ensuremath{L_{\textrm{unstable}}}}$
(seen as a change in slope in figure \ref{fig:FEM-and-local}b at
around $\mbox{<\ensuremath{t_{\textrm{rupt}}}>\ensuremath{\approx}1}$).
For local load sharing, the transition from a metastable to an unstable
network occurs beyond a critical bond-to-bond distance $L_{\textrm{unstable}}$,
whereas for global load sharing this transition occurs beyond a critical
fraction of open bonds and therefore explains the continuous increase
of rupture time as function of time \citep{Bell1978d,Seifert2000,Erdmann2004a,Sprakel2011}.

To test equation \ref{eq:tipping_point}, we show in figure \ref{fig:theory_collapse}b
that all ablation data can be successfully collapsed onto a single
master curve using a normalized ablation size $\frac{l_{\textrm{ablate}}}{L_{\textrm{unstable}}}$.
To compare equation \ref{eq:tipping_point} with both the ablation
data and the typical length scale observed in figure \ref{fig:FEM-and-local}b,
we first define a critical ablation length, $l_{\textrm{crit}}$,
which we obtain by fitting the size dependence of the rupture probability
$\phi_{\textrm{rupt}}$ to a sigmoidal function $\mbox{\ensuremath{\phi_{\textrm{rupt}}}=\ensuremath{\frac{1}{1+e^{-l_{\textrm{crit}}-l_{\textrm{ablate}}}}}}$
at each applied stress and at bond affinity $K=0.9$. We can now combine
the critical ablation length $l_{\textrm{crit}}$ from figure \ref{fig:Characterization-of-hidden}a
with the optimal system size $N$ found in figure \ref{fig:FEM-and-local}c
and conclude that all these data are well-described by equation \ref{eq:tipping_point}
(figure \ref{fig:theory_collapse}c). 

Up to now, theoretical work on transient network fracturing has been
limited to the assumption of global load sharing \citep{Bell1978d,Seifert2000,Erdmann2004a,Sprakel2011}.
We find that $<t_{\textrm{rupt}}>$ is insensitive to this assumption
for microscopic systems (up to approximately 10-100 bonds, see equation
\ref{eq:tipping_point} and figure \ref{fig:FEM-and-local}c). However,
fracturing of larger system follows fundamentally different rules,
in which the notion of local load sharing becomes important. Our study
investigates the idealized limit of fully localized load sharing,
but we observe similar behavior of a typical fracture length when
the load distribution is simulated via a mechanical model \citep{SI}.

Our model predicts features which are different from global load sharing,
but consistent with experimental observations on a wide range of viscoelastic
materials. First, rupturing of macroscopic viscoelastic materials
proceeds via spontaneous crack initiation, different from the homogeneous
failure predicted by global load sharing models \citep{Bell1978d,Seifert2000,Erdmann2004a,Sprakel2011}.
This prediction is borne out by experimental observations of a wide
range of viscoelastic materials \citep{Gladden2007,ABR,Jasper,Foyart2016}.
Second, the model of global load sharing predicts that the rupture
time strongly increases with the system size. As a result, delayed
fracturing ($\mbox{<\ensuremath{t_{\textrm{rupt}}}>\ensuremath{\gg k_{\textrm{on}}}}$)
would only be experimentally observable very close to the critical
stress for any macroscopic system. Instead, delayed fracture is experimentally
observed for many different viscoelastic materials over a wide range
of stresses \citep{Gibaud2010,Sprakel2011,telechelic,Jasper}. We
find that in case of local load sharing, the dependence of $<t_{\textrm{rupt}}>$
on $\sigma$ does not diverge upon increasing system size $N$. Thus,
our model for the first time explains why delayed fracturing is readily
observable on laboratory timescales over a wide range of stresses
in experiments.

The model makes several concrete predictions that can be tested experimentally
by applying shear stress on viscoelastic materials. Firstly, we predict
that the average rupture time measured at constant stress will be
inversely proportional to the system size (figure \ref{fig:FEM-and-local}c)
as the crack initiation rate is constant per volume. Secondly, the
presence of a critical crack length can be measured directly by performing
laser ablation on viscoelastic materials under stress, a technique
that is common in biophysical studies of cell and tissue tension \citep{Tinevez2009}.
Thirdly, the dependence of the critical crack length on the applied
stress and bond kinetics (Equation \ref{eq:tipping_point}) can be
tested experimentally. The bond kinetics can for instance be experimentally
controlled by changing the temperature in cross linked actin networks
\citep{Ward2008} or salt conditions in polyelectrolyte gels \citep{Spruijt2010}.

Our framework for understanding the crack initiation process in viscoelastic
materials can be used to rationally design more robust materials.
We have considered evenly distributed bonds. For future work, it would
be interesting to investigate the effect of inhomogeneity under local
load sharing. It is interesting to note that cellular adhesion proteins
are not randomly distributed but clustered with a well-defined size
\citep{Schwarz2013}. Simulations have shown that an intermediate
degree of clustering is optimal for preventing fracturing \citep{Lin2008,Qian2008},
although the nature of this optimum remained poorly understood. We
speculate that this optimal clustering density is related to the critical
length scale for crack initiation and that this strategy of clustering
bonds is an interesting design principle for synthetic materials.

\section*{Acknowledgements}

We thank Pieter Rein ten Wolde, Chase Broedersz, David Brueckner and
Mareike Berger for fruitful discussions. This work is part of the
research program of the Netherlands Organisation for Scientific Research
(NWO). We gratefully acknowledge financial support from an ERC Starting
Grant (335672-MINICELL).

\bibliographystyle{unsrt}

\part*{{\normalsize{}Supplementary Information}}

\begin{figure}
\includegraphics[width=1\columnwidth]{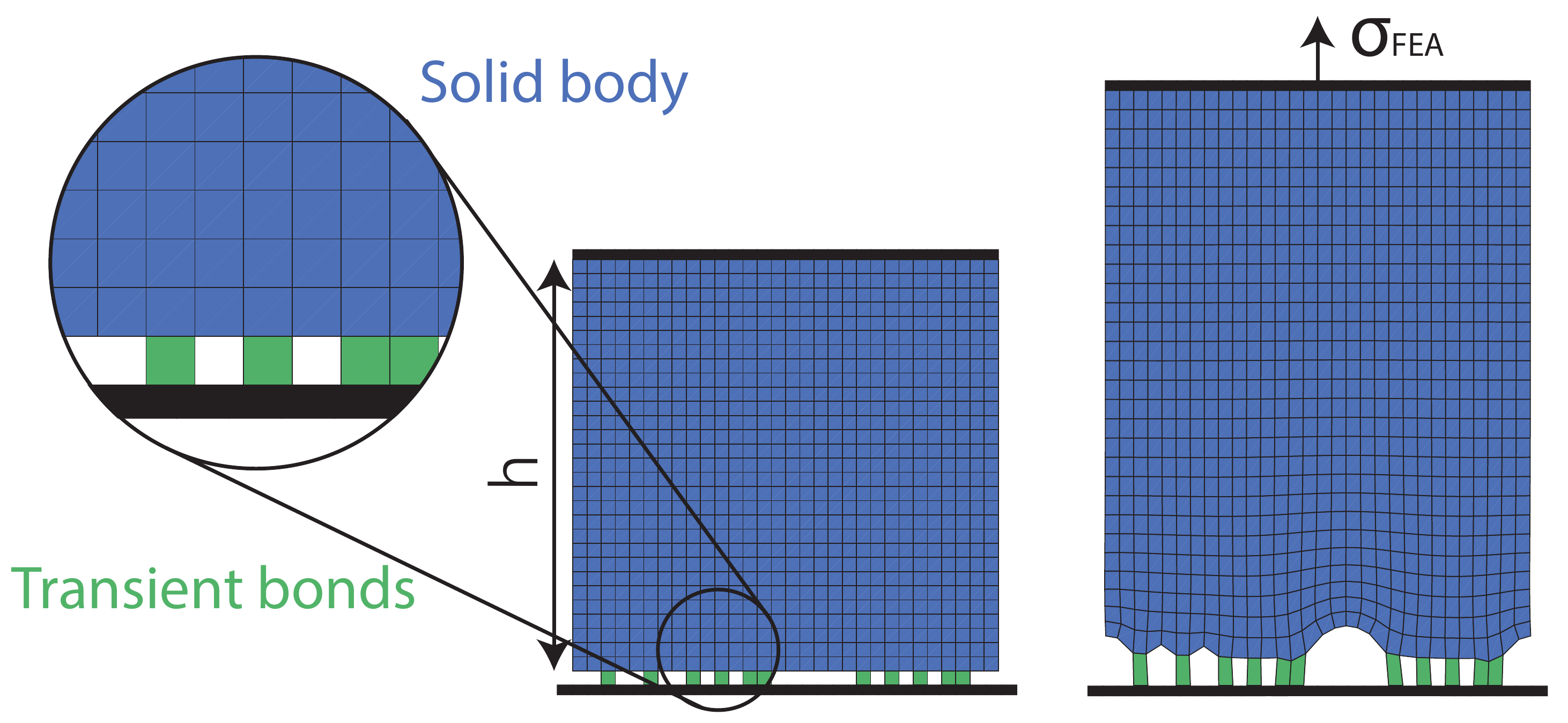}\caption{\textbf{Finite element model }Finite element model representation
of an elastic body (blue) of total thickness $h$ with random distribution
of bonds (green) at rest (left) and stretched (right). The deformation,
due to the applied force $\sigma_{\textrm{FEA}}$ on the elastic body,
is exaggerated for clarity\label{fig:Model-sketch}.}
\end{figure}

\begin{figure}
\includegraphics[width=1\columnwidth]{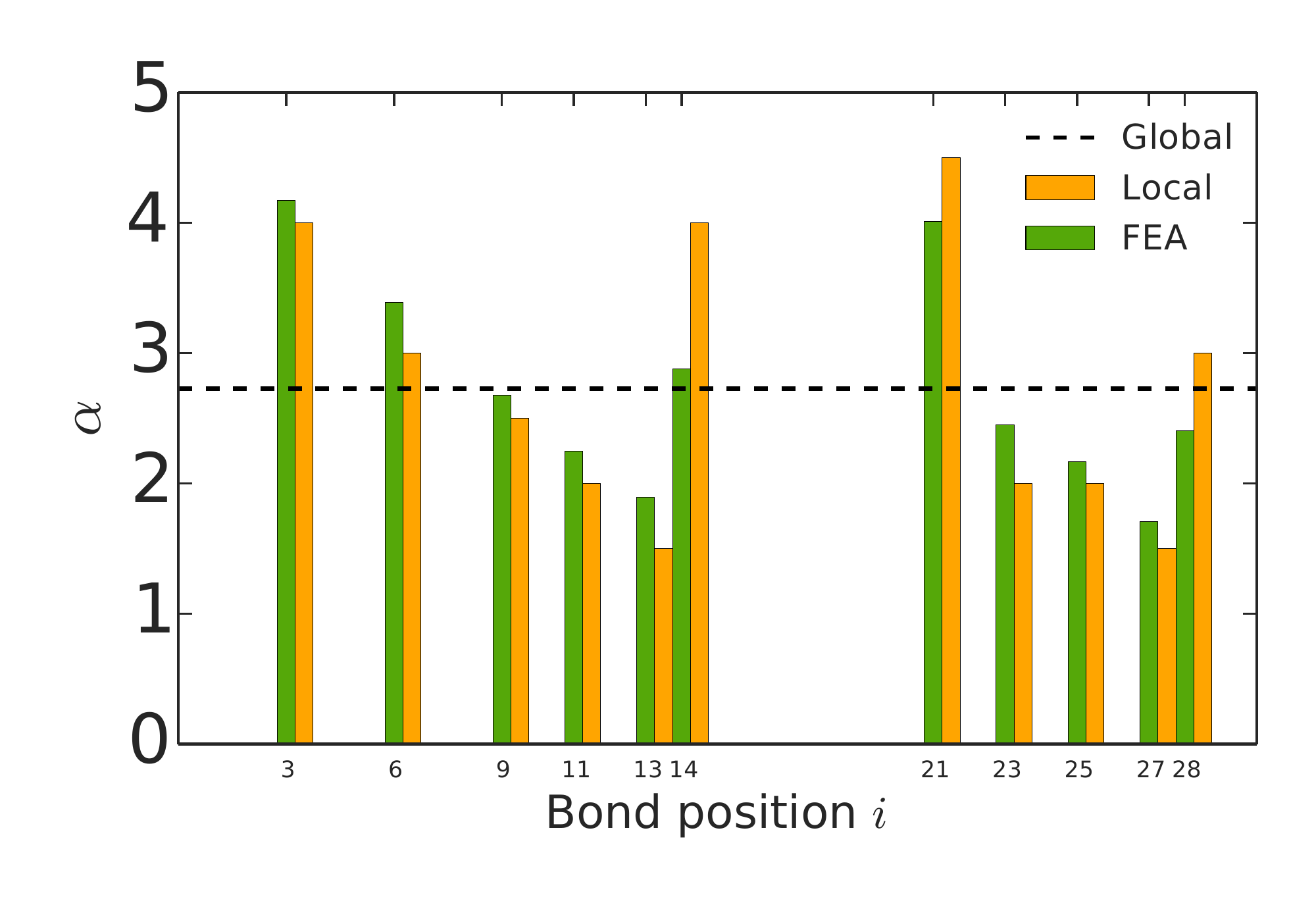}\caption{\textbf{Comparison of load distribution }Stress intensity factor comparison
of different modes of load sharing (global, local, FEA) for the bond
distribution shown in Figure S\ref{fig:Model-sketch}.\label{fig:stress_intensity}.}
\end{figure}

To validate the behavior of cracks in viscoelastic materials obtained
with the local load sharing assumption, we implemented a 2D mechanical
model (figure S\ref{fig:Model-sketch}) based on Finite Element Analysis
(FEA) \citep{Cook2007}. The transient bonds, modeled by a linear
elastic material with a Young's modulus $E$ and Poisson's ratio $\nu$,
are modeled as elastic bodies fixed at their bottom and attached to
an upper body with the same material properties. Both the bonds and
the elastic body are discretized using bi-linear square elements consisting
of four nodes. The elastic body consists of $h\cdot N$ elements,
where $h$ is the height of the elastic body expressed in terms of
the in number of elements and $N$ is the number of bonds. Periodic
boundary conditions are applied to the left and right boundaries.
To apply tension to the bonds, a vertical displacement is applied
to the upper boundary of the solid part, until a force $\sigma_{\textrm{FEA}}$
is reached. A Monte Carlo scheme similar to the one used for the local
load sharing model is applied to determine the transient behavior
of the bonds, using equation 2 and 3 from the main text, and we define
the stress intensity factor value $\alpha_{i}$ for FEA as:
\begin{equation}
\alpha_{i}=\frac{U_{i}}{\sum_{i}U_{i}}N
\end{equation}
where $U_{i}$ is the bonds' elastic strain energy density which can
be found by integrating the stress vector on the bond $\{\boldsymbol{s}\}_{i}$
for the strain vector $\{\boldsymbol{e}\}_{i}$ according to:

\begin{equation}
U_{i}=\int\{\boldsymbol{s}\}_{i}^{T}\{d\boldsymbol{e}\}_{i}=\frac{1}{2}{\left[s_{xx}\quad s_{yy}\quad s_{xy}\right]_{i}\left[\begin{array}{c}
\ensuremath{e_{xx}}\\
\ensuremath{e_{yy}}\\
\ensuremath{e_{xy}}
\end{array}\right]_{i}}
\end{equation}
Figure S\ref{fig:stress_intensity} shows how the elastic body redistributes
the applied force on the bonds. Importantly, we observe a stress intensity
distribution for FEA that is comparable to that of the local load
sharing assumption using the settings of $h=N$ and $E_{body}=E_{bond}$.
For future work it will be interesting to vary $h$ and/or $E_{bond}/E_{body}$
and test the effect on the stress distribution and subsequent fracturing
behavior.

\begin{figure}
\includegraphics[width=1\columnwidth]{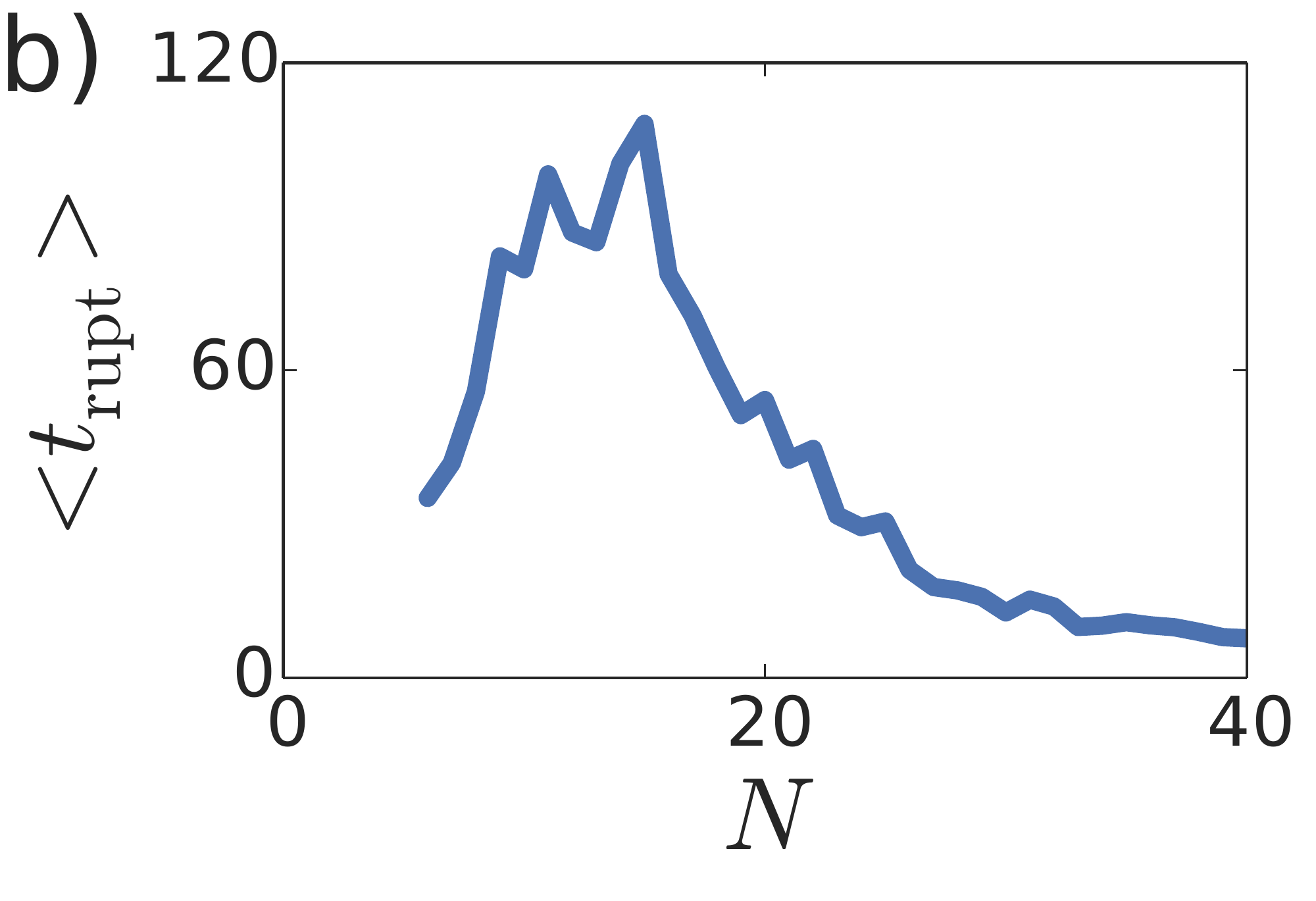}\caption{\textbf{FEA verifies main qualitative difference between local and
global load sharing} Unlike global load sharing but similar to local
load sharing (main text figure 2c),\textbf{ }the average rupture time
in FEA is peaked for intermediate system size $N$.}
\end{figure}
\begin{figure}
\includegraphics[width=1\columnwidth]{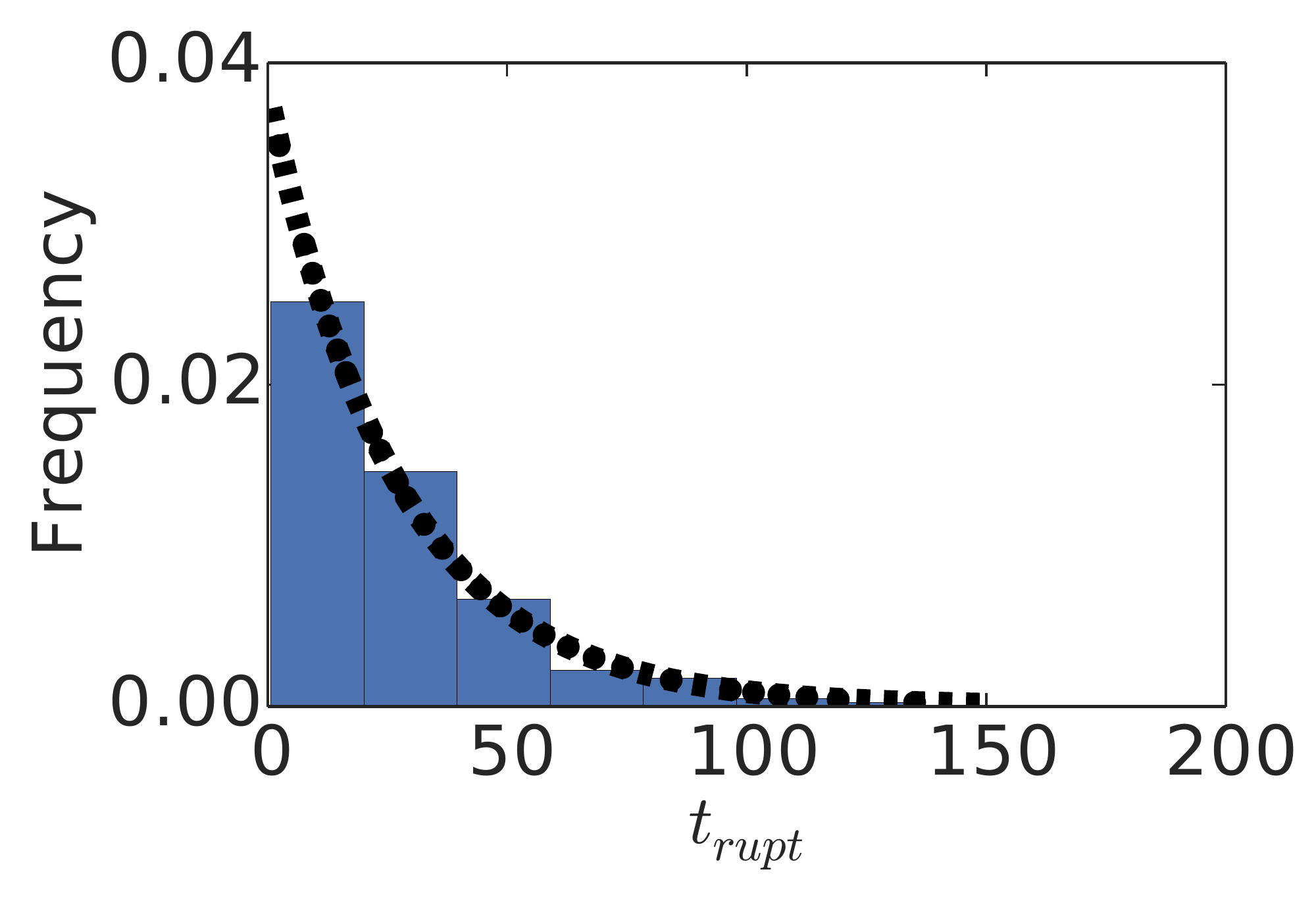}\caption{\textbf{Rupture times are exponentially distributed} We performed
fracturing simulations under local load sharing for 1D networks (see
main text figure 1) using identical parameters ( $\sigma=0.7$, $K=0.9$,
$N=20$, $1000$ repeats) and recorded $\tau_{\textrm{rupt}}$. The
distribution of rupture times is exponential, suggesting a stochastic
process.}
\end{figure}

\end{document}